\documentclass[conference,10pt]{IEEEtran}
\usepackage{verbatim}
\usepackage{varioref}
\usepackage{amsmath}
\usepackage{amssymb}
\usepackage{graphicx}
\usepackage{algorithm}
\usepackage{algpseudocode}
\usepackage{pifont}
\usepackage{sidecap}
\usepackage{acronym}
\usepackage{url}
\usepackage{caption}
\usepackage{subcaption}
\usepackage{color}
\usepackage{stmaryrd}
\usepackage{float}

\makeatletter
\def\BState{\State\hskip-\ALG@thistlm}
\makeatother

\usepackage{cite}
\acrodef{CS}[CS]{compressed sensing}
\acrodef{AWGN}[AWGN]{additive white Gaussian noise}
\acrodef{MSE}[MSE]{mean squared error}
\acrodef{GAMP}[GAMP]{generalized approximate message passing}
\acrodef{SNR}[SNR]{signal-to-noise ratio}
\acrodef{SR}[SR]{self-reset}
\acrodef{pdf}[pdf]{probability density function}
\acrodef{ADC}[ADC]{analog to digital converter}
\acrodef{MC}[MC]{Monte-Carlo}
\makeatother

\newcommand{\mbsf}[1]{\boldsymbol{\mathsf{#1}}}

\begin{document}

\title{Generalized Approximate Message Passing for Unlimited Sampling of Sparse Signals}

\author{$\text{Osman Musa}^{\dagger\star}$, $\text{Peter Jung}^{\dagger}$ and $\text{Norbert Goertz}^{\star}$  \\
$\text{}^{\dagger}$Communications and Information Theory, Technische Universit{\"a}t Berlin\\
$\text{}^{\star}$Institute of Telecommunications, Technische Universit{\"a}t Wien\\
Email: \{osman.musa,norbert.goertz\}@nt.tuwien.ac.at, peter.jung@tu-berlin.de}
\maketitle

\begin{abstract}
In this paper we consider the \ac{GAMP} algorithm for recovering a sparse signal from modulo samples of randomized projections of the unknown signal. The modulo samples are obtained by a \ac{SR} \ac{ADC}. Additionally, in contrast to previous work on \ac{SR} \ac{ADC}, we consider a scenario where the \ac{CS} measurements (i.e., randomized projections) are sent through a communication channel, namely an \ac{AWGN} channel before being quantized by a \ac{SR} \ac{ADC}. To show the effectiveness of the proposed approach, we conduct \ac{MC} simulations for both noiseless and noisy case. The results show strong ability of the proposed algorithm to fight the nonlinearity of the \ac{SR} \ac{ADC}, as well as the possible additional distortion introduced by the  \ac{AWGN} channel.
\end{abstract}

\begin{IEEEkeywords}
Generalized approximate message passing, self-reset analog to digital converter, noisy channel, compressed sensing, Bernoulli-Gaussian mixture
\end{IEEEkeywords}

\vspace*{-4mm}

\begin{figure*}[h]
\hspace*{-1.2cm}
\begin{centering}
\includegraphics[scale=0.5]{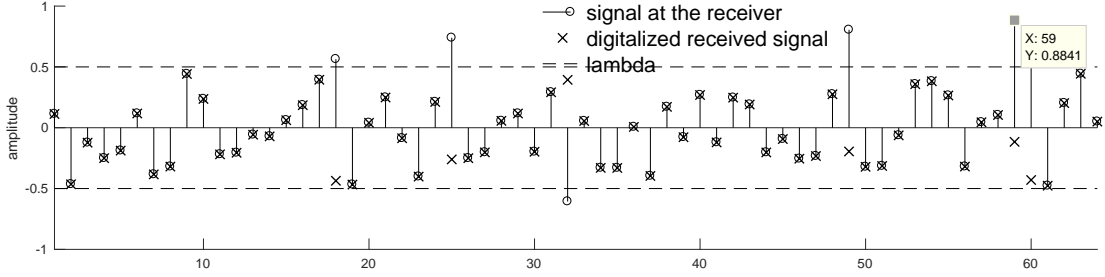}
\par\end{centering}
\vspace{-0ex} \protect\caption{An example of digitalizing a signal with \ac{SR} \ac{ADC}, with $\lambda = 0.5$. All the values inside interval $[-\lambda, \lambda]$ are kept undistorted, while the values outside this range are folded back to the interval $[-\lambda, \lambda]$. 
\label{fig:sradc_example}}
\vspace{-2ex}
\end{figure*}

\acresetall
\section{Introduction}

Whittaker-Nyquist-Kotelnikov-Shannon theorem is the fundamental result in signal processing, that states that one can perfectly reconstruct a continuous bandlimited signal from a set of samples, taken at a sampling rate which is proportional to bandwidth of the signal. Here we assume that the \ac{ADC} has infinite precision and infinite dynamic range. Even though, the theory of finite precision quantization (rate distortion theory) is well known for decades, the effects of finite dynamic range (i.e., clipping) became interesting only recently in different research communities, e.g., in image and audio processing, bio-medical applications and analysis of physiological data \cite{restoring_clipped_signal,Aliasing_Reduction_in_Clipped_Signals,audio_declipping}.  

To reduce the negative effects of clipping, Bhandari \textit{et al.} \cite{on_ULS} propose digitalizing bandlimited signals with a \ac{SR} \ac{ADC} with an appropriate choice of the threshold parameter $\lambda$. 
The \ac{SR} \ac{ADC} with the parameter $\lambda$ is defined by the mapping 
\begin{equation}
\label{eq:sradc}
\mathcal{M}_{\lambda}(t) = 2\lambda \bigg( \bigg\llbracket \frac{t}{2\lambda} + \frac{1}{2} \bigg\rrbracket  - \frac{1}{2}\bigg) ,
\end{equation}
where $\llbracket t \rrbracket \triangleq t- \lfloor t \rfloor$ is the remainder of the division $t$ and $\lambda$.

In Fig. \ref{fig:sradc_example} we illustrate the effects of digitalization with \ac{SR} \ac{ADC}, where one can observe that only values of the received signal that are outside the range $[-\lambda, \lambda]$ are affected by the \ac{ADC} in the sense that the input value is folded to the range $[-\lambda, \lambda]$.
If some estimate of the norm of the input signal is known, the authors of \cite{on_ULS} prove that perfect recovery of a bandlimited signal from its discrete samples is possible if the sampling period $T \leq (2\pi e)^{-1}$, where it is assumed that the bandwidth of the signal is normalized to $\pi$. Apart from giving the sufficient conditions for perfect recovery, the authors present a stable recovery algorithm.  

When sampling certain sparse signals, it was reported in \cite{ULSSS, FRI_sampling_theory, SD_UWB_Receiver, gp_radar}, that  during the calibration phase, the received amplitudes are typically larger than during the subsequent sensing phase. Unlike classical approaches of clipping or saturation, the authors in \cite{ULSSS}  provide sufficient conditions for perfect recovery of $K$-sparse\footnote{A $K$-sparse vector has at most $K$ nonzero components} signal from its low-pass filtered version, together with a constructive recovery algorithm.

\subsection*{Contributions}
\begin{figure*}
\begin{centering}
\includegraphics[scale=0.75]{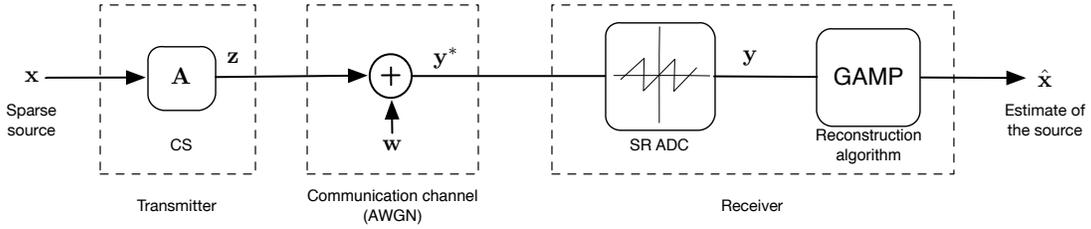}
\par\end{centering}
\protect\caption{The signal processing chain. The unknown $K$-sparse vector $\mathbf x \in \mathbb{R}^N$ is multiplied with measurement matrix $\mathbf A_{n\times N}$ to obtain a vector of CS measurements $\mathbf z \in \mathbb{R}^n$. The components of $\mathbf z$ are transmitted through an \ac{AWGN} channel. At the receiver, the samples of the received signal $\mathbf y^{\ast}$ are digitalized with a \ac{SR} \ac{ADC} to obtain the vector of measurements $\mathbf y$. The \ac{GAMP} algorithm is applied to produce an estimate $\hat{\mathbf x}$ of the unknown sparse signal $\mathbf x$.
\label{fig:sheme}}
\vspace{-2ex}
\end{figure*}

In this paper we follow the work of \cite{ULSSS}, but instead of sampling a low-pass filtered version of a sparse signal, we take \ac{CS} measurements and digitalize them with a \ac{SR} \ac{ADC}. This problem corresponds to the communication scenario shown in Fig. \ref{fig:sheme}, where we first construct a vector of \ac{CS} measurements of a sparse signal. That message vector is later transmitted through an \ac{AWGN} channel and digitalized at the receiver with a \ac{SR} \ac{ADC}. To recover the unknown sparse signal we employ the well known \ac{GAMP} \cite{GAMP} algorithm and tailor it to our specific problem. The \ac{GAMP} algorithm was already successfully applied in \cite{GAMP, 1BitGAMP, BiLiClass, Message-Passing_De-Quantization_With_Applications_to_Compressed_Sensing, noisy_QCS_GAMP, GAMP_for1bit_AWGN} for recovery of sparse signals from \ac{CS} measurements with nonlinear distortions. To our best knowledge this is the first work that examines the effects of \ac{SR} \ac{ADC} on \ac{CS} phase transition curves.

\subsection*{Notation} 
Vectors and matrices are represented by boldface characters. 
Random variables, random vectors, and random matrices are denoted by sans-serif font, e.g., $\sf a$, $\mbsf{a}$, and $\mbsf{A}$, respectively. Function $n(x; \mu, \sigma^2)$ represents a Gaussian pdf with mean $\mu$ and variance $\sigma^2$ evaluated at $x$. The Hadamard product (i.e., component-wise multiplication) is denoted by the operator $\bullet$. If a scalar valued function receives a vector as its argument, this means component-wise application of that function. For example, $\mathcal{M}(\mathbf z) = [\mathcal{M}(z_1),...,\mathcal{M}(z_n)]^T$, and $(\mathbf z)^{-1} = [z_1^{-1},...,z_n^{-1}]^T$. The Dirac delta distribution is represented by $\delta (\cdot)$. Unless otherwise specified $\|\cdot\|$ corresponds to the Euclidian ($l_2$) norm.

\section{Self Reset Analog to Digital Conversion of CS Measurements Corrupted with \ac{AWGN}}
Next, we formulate the mathematical model for the unknown signal and the measurement process.

\subsection{Signal Model}

We assume that the components $\{{\sf x}_i \}_{i=1}^N$ of the unknown sparse vector $\mbsf{x}$ are i.i.d.~realizations of the Bernoulli-Gaussian mixture distribution, i.e., 
\begin{equation}
\label{eq:px}
p_{{\sf x}_i}(x) = (1-\epsilon) \delta (x) + \epsilon \, n(x; 0, \sigma^2),
\end{equation}
where $\epsilon$ represents the probability of nonzero value. Consequently, $1- \epsilon$ is the sparsity of the signal.

\subsection{Measurement Model}
Each measurement $\sf{y}_i$ is a folded version of the corresponding component $i$ of the received signal $\mbsf{y}^{\ast}$, i.e.,
\begin{equation}
\sf{y}_i = \mathcal{M}_{\lambda} (\sf{y}^{\ast}_i),
\end{equation}
where $\mathcal{M}_{\lambda}(\cdot)$ represents the nonlinear mapping of the \ac{SR} \ac{ADC} converter given in (\ref{eq:sradc}). We note that the involved \ac{SR} \ac{ADC} has infinite precision in the interval $[-\lambda, \lambda]$. Alternatively, respecting (\ref{eq:sradc}) we can write 
\begin{equation}
\mbsf{y}^{\ast} = \mbsf{y} + \mbsf{\epsilon}_g,
\end{equation}
where the entries of vector $\mbsf{\epsilon}_g$ are samples so-called simple function. These samples belong to a set of discrete points $2 \, \lambda \, \mathbb Z$.
Furthermore, $\mbsf{y}^{\ast}$ is equal to the sum of the vector of \ac{CS} measurements $\mbsf{z}$ and a noise vector $\mbsf{w}$, i.e.,
\begin{equation}
\mbsf{y}^{\ast} = \mbsf{z} + \mbsf{w} = \mbsf{A} \mbsf{x} + \mbsf{w},
\end{equation}
where $\mbsf{w}$ is i.i.d. zero-mean \ac{AWGN} noise vector with the covariance matrix $\sigma_w^2 \, \mathbf I$, i.e., $\mbsf{w} \sim \mathcal{N}(\mathbf 0,\sigma_w^2 \, \mathbf I)$, and $\mbsf{A} \in \mathbb R^{n \times N}$ is the Gaussian measurement matrix, that defines the sampling rate (indeterminacy) $\rho = n/N$. Finally, we can compactly write
\begin{equation}
\label{eq:measurement_model}
\mbsf{y}= \mathcal{M}_{\lambda} \big(\mbsf{A} \mbsf{x} + \mbsf{w}\big).
\end{equation}

Our goal is to estimate $\mathbf x$ from $\mathbf y$. To solve this problem we employ the GAMP algorithm that we present in the next section.

\section{The Generalized Approximate Message Passing Algorithm or Self-Reset \ac{ADC}}

\subsection{The GAMP Algorithm}

\begin{table*}
\begin{center}
\vspace{2mm}
\begin{tabular}{ |p{1.2cm}||p{5cm}|p{5cm}|p{4cm}|  }
\hline
& \parbox{5cm}{ \centering \begin{equation*} \mathbb{E} \{ \mathsf z \mid \mathsf y =  y  \}  		\end{equation*}} 
&  \parbox{5cm}{ \centering \begin{equation*} \mathbb{E} \{ \mathsf z^2 \mid \mathsf y =  y \} 		\end{equation*} } 
& \parbox{4cm}{\centering \begin{equation*} f_\mathsf y(y) \end{equation*}}   \\
\hline
\hline
\ac{SR} \ac{ADC}  
& \parbox{5cm}{ \centering \begin{equation*} \frac{1}{f_\mathsf y(y)} \sum_{k = -\infty}^{\infty} (y+2k\lambda) \thinspace n(y+2k\lambda;\mu_z, \sigma^2_z) 		\end{equation*}} 
& \parbox{5cm}{ \centering \begin{equation*} \frac{1}{f_\mathsf y(y)} \sum_{k = -\infty}^{\infty} (y+2k\lambda)^2 \thinspace n(y+2k\lambda;\mu_z, \sigma^2_z) 	\end{equation*}} 
& \parbox{4cm}{ \centering \begin{equation*} \sum_{k = -\infty}^{\infty} n(y+2k\lambda;\mu_z, \sigma^2_z) 						\end{equation*}}   \\
\hline
\ac{AWGN} 
& \parbox{5cm}{ \centering \begin{equation*} \bigg(\frac{y}{\sigma_w^2} + \frac{\mu_z}{\sigma^2_z}\bigg) \sigma^2_{wz}		\end{equation*}} 
& \parbox{5cm}{ \centering \begin{equation*}\frac{\sigma_w^2\sigma_z^2}{\sigma_w^2 + \sigma_z^2}  + \mathbb{E} \{ \mathsf z \mid \mathsf y =  y  \}^2	\end{equation*}}  
& \parbox{4cm}{ \centering \begin{equation*} \sum_k n(0; y + 2k\lambda - \mu_z,  \sigma^2_z +  \sigma^2_w)				\end{equation*}}    \\
\hline
\end{tabular}
\vspace{1mm}
\caption{Scalar mean, power, and probability density function for the GAMP nonlinear measurement updates.} 
\vspace{-6mm}
\end{center}
\end{table*}

\label{algorithm}
The equations (\ref{eq:t0}-\ref{eq:estimate}) define the steps of the \ac{GAMP} algorithms \cite{GAMP}.
\begin{enumerate}
\item \textit{Initialization:} At $t=0$, respecting the prior in (\ref{eq:px}), the \ac{GAMP} algorithm is intialized according to
\begin{equation}
\label{eq:t0}
\mathbf{\hat{\mathbf  x}}^{0}=\mathbb{E} \{\mathbf  x\} =0, \enspace \mathbf{v}^{0}_{\mathbf  x}  = \text{var}\{\mathbf  x\} = \epsilon \,\sigma^2, \enspace \mathbf{\hat{\mathbf  s}}^{0} =\mathbf  y.
\end{equation}


\item \textit{Iteration:} At every subsequent iteration $t=1,2,...,t_{\text{max}}$ it performs the measurement updates before the estimation updates, where both updates have a linear step followed by a nonlinear step. Those updates are calculated according to:
\begin{enumerate}
\item \textit{Measurement update linear step:} 
\begin{subequations}
\begin{alignat}{1}
\mathbf{v}^{t}_p & = (\mathbf{A} \bullet \mathbf{A}) \mathbf{v}^{t-1}_x, \\
\mathbf{\hat{p}}^{t} &= \mathbf{A} \mathbf{\hat{x}}^{t-1} - \mathbf{v}^{t}_p \bullet \mathbf{\hat{s}}^{t-1}.
\end{alignat}
\end{subequations}
\item \textit{Measurement update nonlinear step:}
\begin{subequations}
\label{eq:F}
\begin{alignat}{1}
\mathbf{\hat{s}}^{t}  & = \text{F}_1(\mathbf{y}, \mathbf{\hat{p}}^{t} , \mathbf{v}^{t}_p), \\
\mathbf{v}^{t}_s &= \text{F}_2(\mathbf{y}, \mathbf{\hat{p}}^{t} , \mathbf{v}^{t}_p).
\end{alignat}
\end{subequations}
\item \textit{Estimation update linear step:} 
\begin{subequations}
\begin{alignat}{1}
\label{vt_r}
\mathbf{v}^{t}_r & = \big((\mathbf{A} \bullet \mathbf{A})^T \mathbf{v}^{t}_s\big)^{-1}, \\
\mathbf{\hat{r}}^{t} &= \mathbf{\hat{x}}^{t-1} + \mathbf{v}^{t}_r \bullet (\mathbf{A}^T \mathbf{\hat{s}}^{t}).
\end{alignat}
\end{subequations}
\item \textit{Estimation update nonlinear step:} 
\begin{subequations}
\label{eq:estimate}
\begin{alignat}{1}
\mathbf{\hat{x}}^{t}  & = \text{G}_1(\mathbf{\hat{r}}^{t}, \mathbf{v}^{t}_r,p_x), \\
\mathbf{v}^{t}_x &= \text{G}_2(\mathbf{\hat{r}}^{t}, \mathbf{v}^{t}_r,p_x).
\end{alignat}
\end{subequations}
\end{enumerate}

The nonlinear functions in (\ref{eq:F}) and (\ref{eq:estimate}) are applied component-wise and are given by
\begin{equation}
\label{FandG}
\begin{alignedat}{4}
\text{F}_1(y, \hat{p} , {v}_p)  & \,{=}\, \frac{\mathbb{E} \{ \mathsf z \vert \mathsf y\} - \hat{p}}{v_p}, \enspace & \text{G}_1(\hat{r}, v_r,p_x)  & \,{=}\, \mathbb{E} \{ \mathsf x \vert \hat{\mathsf r} \},\\
\text{F}_2(y, \hat{p} , {v}_p) & \,{=}\, \frac{v_p - \text{var} \{ \mathsf z \vert \mathsf y\} }{v^2_p}, \> & \text{G}_2(\hat{r}, v_r,p_x) & \,{=}\, \text{var}\{\mathsf x \vert \hat{\mathsf r} \},
\end{alignedat}
\end{equation}
where
\begin{equation}
\begin{alignedat}{2}
f_{\mathsf z \vert \mathsf y} & \propto f_{\mathsf y \vert \mathsf z} \, f_{\mathsf z}  = f_{\mathsf y \vert \mathsf z} \enspace n(\cdot;\hat p, v_p), \\
f_{\mathsf x \mid \hat{\mathsf r}} & \propto f_{\hat{\mathsf r} \mid\mathsf x}  \, f_\mathsf x  =  n(\cdot;\hat r, v_r) \, f_\mathsf x.
\end{alignedat}
\end{equation} 
\setcounter{enumi}{2}
\item \textit{Stopping criterion:} We define two criteria for the determining the convergence of the algorithm. We stops iterating if ${\lVert \mathbf{\hat x}^t - \mathbf{\hat  x}^{t-1}\rVert_2 < \varepsilon \: \lVert \mathbf{\hat x}^t \rVert_2}$
with a small $\varepsilon>0$ (e.g.,~$\varepsilon = 10^{-2}$) or when $t \geq t_{\text{max}}$, where $t_{\text{max}}$ is predefined maximum number of iterations (typically in the order of $N$ or less).
\end{enumerate}

To get more accurate estimate, we use the vector version of the algorithm. Therefore we do not average over the entries of $\mathbf{v}^{t}_s$  and $\mathbf{v}^{t}_x$, given in (\ref{eq:F}) and (\ref{eq:estimate}), respectively.

\subsection{Nonlinear Steps in the Updates}
Given the fact that $\mathsf z \sim \mathcal{N} (\mu_z, \sigma_z^2)$\footnote{Here we use $\mu_z$ and $\sigma_z^2$, instead of $\hat p$ and $v_p$, respectively} and considering the measurement model given by (\ref{eq:measurement_model}), we can calculate the closed form expressions for the scalar measurement updates in (\ref{FandG}). These terms are computed according to Table 1.

The expressions for the nonlinear functions $\text{G}_1(\cdot)$ and $\text{G}_2(\cdot)$ are identical to those in \cite{noisy_QCS_GAMP}.

\section{Numerical Results}
To investigate the performance of the proposed reconstruction algorithm we perform \ac{MC} simulations, with the associated parameters described in the following subsection.

\subsection{Simulation Setup}

The measurement ratio $\rho$ and the probability of nonzero value $\epsilon$ take values in the range $[0.1, 1]$ and $[0.0156, 0.25]$, respectively. For a specific pair $\{\rho,\epsilon\}$, we average results over $4000$ independent realizations of sets indices of
nonzero components, the values of the nonzero components, the Gaussian sensing
matrix $\mathbf A$, and the AWGN $\mathbf  w$.  The nonzero components of the
source vector $\mathbf x$ as well as the entries of the measurement matrix
are drawn randomly from a zero-mean Gaussian distribution with power $\sigma^2=1$ and $\sigma^2=1/n$, respectively.  
In each simulation we fix $N=256$, and acquire $n=\rho\,N$ measurements of the $K=\epsilon\,N$ sparse vector. 
Each CS measurement vector is corrupted with AWGN noise with power $\sigma_w^2 = 10^{-\text{SNR}/10}$, where the SNR is defined as
\begin{equation}
\text{SNR}/\text{dB} = 10 \log_{10}\{\lVert \mathbf y^{\ast} \lVert^2 / \lVert \mathbf w\lVert^{2} \}. 
\end{equation}
In the noiseless case, we simply set $\text{SNR} = \infty$. The \ac{SR} \ac{ADC} threshold $\lambda$ is fixed to $1$.

The stopping threshold for the algorithms is
$\varepsilon=10^{-3}$, where as the maximal number of iterations of the
proposed algorithm is set to $t_{\text{max}} = N/2 = 128$.

To get an insight at recovery potential of the \ac{GAMP} algorithm, we calculate \ac{MSE} for each independent realization of $\mathbf x$, which is defined as
\begin{equation}
\label{eq:MSE}
\text{MSE} / \text{dB} =10 \log_{10}  \lVert \mathbf x - \hat{ \mathbf x} \lVert^2_2.
\end{equation}
In the noiseless case, we calculate the success rate as the average number of successful recoveries. A recovery is considered successful if the resulting \ac{MSE} is $\leq -30\text{dB}$. We chose this measure of quality of the reconstruction since in the noiseless case, the algorithm either recovers the unknown signal almost perfectly (with very small \ac{MSE} $\leq$ - 40dB), or fails completely. In the noisy case, \ac{MSE} is used as a figure of merit.

\begin{figure*}
\vspace{-4mm}
\centering
\begin{subfigure}{.5\textwidth}
  \centering
  \includegraphics[scale=0.4]{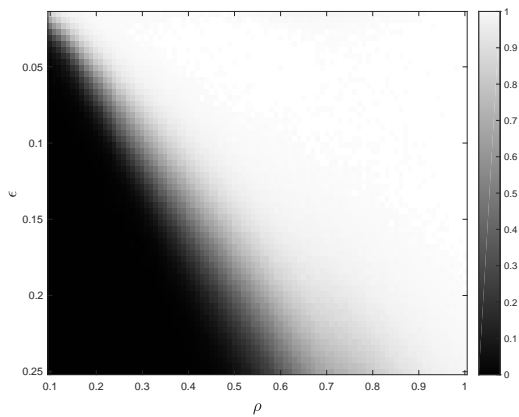}
  \caption{Average success rate of \ac{GAMP}}
  \label{fig:noiseless1}
\end{subfigure}%
\begin{subfigure}{.5\textwidth}
  \centering
  \includegraphics[scale=0.4]{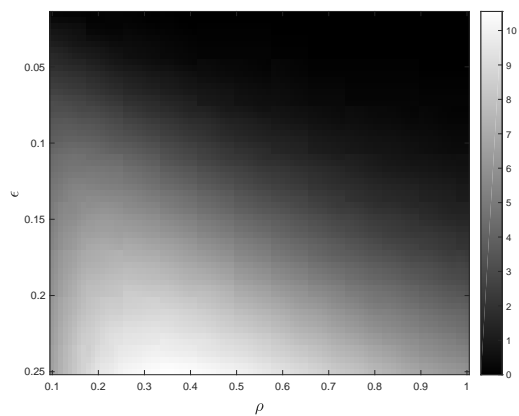}
  \caption{Average norm of the simple function $\lVert \mbsf{\epsilon}_g \lVert_0$}
  \label{fig:noiseless2}
\end{subfigure}
\caption{Average success rate of \ac{GAMP} reconstruction algorithm on the left, and average norm of the simple function $\lVert \mbsf{\epsilon}_g \lVert_0$ on the right as a function of the nonzero probability $\epsilon$ and the measurement ratio $\rho$. The \ac{CS} measurements are digitalized with a \ac{SR} \ac{ADC} with $\lambda=1$. We consider a reconstruction to be successful if the corresponding reconstruction \ac{MSE} is $\leq -30$dB.}
\label{fig:noiseless}
\end{figure*}

\begin{figure*}
\vspace{-4mm}
\centering
\begin{subfigure}{.5\textwidth}
  \centering
  \includegraphics[scale=0.4]{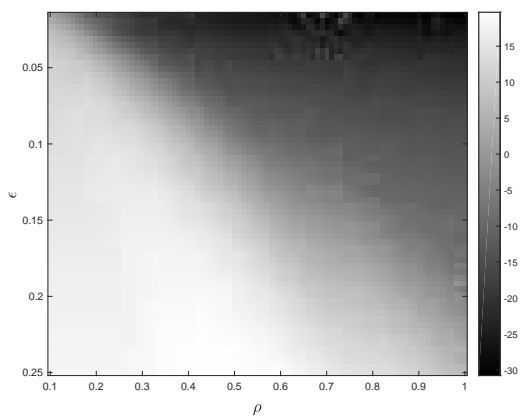}
  \caption{Average \ac{MSE} in dB of \ac{GAMP}}
  \label{fig:noisy1}
\end{subfigure}%
\begin{subfigure}{.5\textwidth}
  \centering
  \includegraphics[scale=0.4]{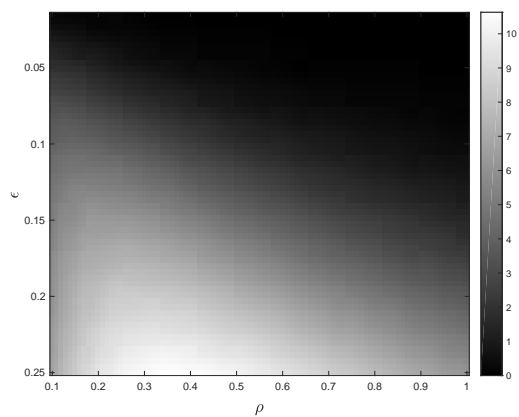}
  \caption{Average norm of the simple function $\lVert \mbsf{\epsilon}_g \lVert_0$}
  \label{fig:noisy2}
\end{subfigure}
\caption{Average \ac{MSE} in dB of \ac{GAMP} reconstruction algorithm on the left, and average norm of the simple function $\lVert \mbsf{\epsilon}_g \lVert_0$ on the right as a function of the nonzero probability $\epsilon$ and the measurement ratio $\rho$. The \ac{CS} measurements are corrupted with \ac{AWGN} noise before being digitalized with a \ac{SR} \ac{ADC} with $\lambda=1$. The \ac{SNR} is set to 20dB.}
\label{fig:noisy}
\vspace{-4mm}
\end{figure*}

\subsection{Results}
\subsubsection*{Noiseless Case} In Fig. \ref{fig:noiseless}, we show the success rate of the \ac{GAMP} algorithm  (Fig. \ref{fig:noiseless1}) and the average norm of the simple function $\lVert \mbsf{\epsilon}_g \lVert_0$ (Fig. \ref{fig:noiseless2}), both as a function of the measurement ratio $\rho$ and the nonzero probability $\epsilon$. The norm of the simple function provides a measure of how corrupted the measurements are due to \ac{SR} \ac{ADC}. In Fig. \ref{fig:noiseless1}, we see a clear phase transition between unsuccessful (black) and successful (white) regions.  While classical \ac{CS} algorithms completely fail when $\lVert \mbsf{\epsilon}_g \lVert_0 \neq 0$, we observe that \ac{GAMP} is able to cope with folded measurements, and the phase transition is almost linear in $\epsilon$. 
\subsubsection*{Noisy Case}  In Fig. \ref{fig:noisy}, we show the \ac{MSE} of the \ac{GAMP} algorithm  (Fig. \ref{fig:noisy1}) and the average norm of the simple function $\lVert \mbsf{\epsilon}_g \lVert_0$ (Fig. \ref{fig:noisy2}), both as a function of the measurement ratio $\rho$ and the nonzero probability $\epsilon$. In Fig. \ref{fig:noisy1} we observe that, compared to the noiseless case, the phase transition curve is shifted the right lower corner. This is to be expected, since the measurements are corrupted with \ac{AWGN} ($\text{SNR}=20$dB) before digitalization, and more measurements are needed for accurate reconstruction.
\subsubsection*{Comments}
It should be noted that if $\lambda \to 0$ the measurements become less and less informative, and in the limit they carry no information. However, taking too large $\lambda$, in practical scenarios with finite bit-budget per sample leads to coarse quantization. Hence, one needs to make a good trade-off between large dynamic range and fine quantization resolution. Therefore, it is an interesting research problem to investigate the effects of folding combined with finite bit budget quantization of the measurements on the \ac{CS} phase transition curves.

\vspace{-3mm}
\section{Conclusions}
In this paper we investigated the potential of applying the \ac{GAMP} algorithm for recovery of sparse signal from \ac{CS} measurements digitalized with a \ac{SR} \ac{ADC}. Our results show that for certain choice of the signal parameters, the \ac{GAMP} algorithm is able to successfully recover a sparse signal from folded measurements. Moreover, unlike the previously proposed algorithm for recovery of sparse signals from folded measurements, the \ac{GAMP} algorithm can cope with the noise introduced by a communication channel.
\bibliographystyle{myieeetran} \bibliography{references}

\end{document}